\documentclass [aps,pre,twocolumn,showpacs] {revtex4}
\usepackage[final]{graphics}
\usepackage{amssymb}
\usepackage{amsfonts}
\usepackage{epsfig}

\begin{document}
\date{\today}
\title{Probability Distribution Function of the Order Parameter: Mixing Fields and Universality}
\author{J. A. Plascak$^{1,2}$}
\email[]{pla@fisica.ufmg.br}
\author{P. H. L. Martins$^{2,3}$}
\email[]{pmartins@fisica.ufmt.br}
\affiliation{$^1$Departamento de F\'\i sica, Instituto de Ci\^encias Exatas,
 Universidade Federal de Minas Gerais, C.P. 702,
 30123-970 Belo Horizonte, MG - Brazil\\
 $^2$Center for Simulational Physics, University of Georgia, Athens, GA
30602, USA \\
 $^3$Instituto de F\'\i sica, Universidade Federal de Mato Grosso,
Av.Fernando Corr\^ea da Costa 2367, 78060-900
 Cuiab\'a, MT - Brazil}

\begin{abstract}

We briefly review the use of the order parameter probability distribution function as a useful tool to obtain the critical properties of statistical mechanical models using computer Monte Carlo simulations. Some simple discrete spin magnetic systems on a lattice, such as Ising, general spin-$S$ Blume-Capel and Baxter-Wu, $Q$-state Potts, among other models, will be considered as examples. The importance and the necessity of the role of mixing fields in asymmetric magnetic models will be discussed in more detail, as well as the corresponding distributions of the extensive conjugate variables.

\end{abstract} 

\maketitle
\section{ Introduction}

According to the words of Sir Humphry Davy (1778-1829)\cite{davy}, {\it ``Nothing tends to the advancement of knowledge as the application of a new instrument."} As a matter of fact, several examples can be traced back to levers, thermometers, telescopes and, perhaps nowadays, to computers. In particular, the use of the thermometer, developed during the 16th and 17th centuries, amongst other fantastic applications, also marked the dawn of the thermal physics and phase transitions era (not counting, of course, the discovery of fire itself). Apart from its rapid technical application in medicine, it is believed that the thermometer in thermal physics can be rightly compared to the use of the telescope in astronomy. It also made possible the discovery of the specific heat and latent heat by Joseph Black (1728-1799) in Scotland (see, for instance, references \cite{breche} and \cite{aps}). On April 1762, Black coined the term {\it latent heat} to refer to the amount of energy required to change the phases solid $\leftrightarrow$ liquid or liquid $\leftrightarrow$ vapor without changing their temperature. This is nowadays known as a first-order, or discontinuous, phase transition. 

In 1822, Charles Cagniard de la Tour (1777-1859) in France discovered {\it critical phenomena} \cite{breche}. Although Cagniard actually saw the liquid and vapor fluid phases becoming equal, the term {\it critical} was only coined some years later, in 1859, by Thomas Andrews (1813-1885) through the observation of what he called {\it critical opalescence}, first observed in carbon dioxide (carbon dioxide was discovered by the same Joseph Black almost one century before). At this critical point there is neither distinction between the liquid and vapor phases nor latent heat, and such a transition is nowadays known as a second-order, or continuous, phase transition (for a recent review on phase transitions see Kadanoff \cite{kadanoff}, and also Domb \cite{domb}).

Together with the advent of these completely new phenomena, a question was raised: the determination of the critical point for different substances. According to the Gibbs phase rule, a single component system can present only isolated critical points, as well as only single triple points \cite{chandler}. Although the critical temperature $T_c$ and the critical pressure $p_c$ are rather easily measured, the measurement of the critical density $\rho_c$ is not so straightforward and turns out to be a much more difficult task. This can be seen in Figure \ref{lrd} from the experimental results of three different compounds.  As an example, in this Figure one has the original data reported by Louis Paul Cailletet (1832-1913) and \'Emile Ovide Joseph Mathias (1861-1942), where the lack of data to measure $\rho_c$ is apparent\cite{cail-mat}. It was actually Cailletet and Mathias who proposed what was later (and still is) known as the law of the rectilinear diameter
\begin{equation}
\rho_d={{1}\over{2}}(\rho_L+\rho_V)=\rho_o+A T,
\label{lrd}
\end{equation}
where $\rho_d$ is the diameter density, $\rho_L$ and $\rho_V$ are the liquid and vapor densities, respectively, and $\rho_o$ and $A$ are constants which depend on the substance. 
%
\begin{figure}[ht]
\includegraphics[clip,angle=0,width=6.5cm]{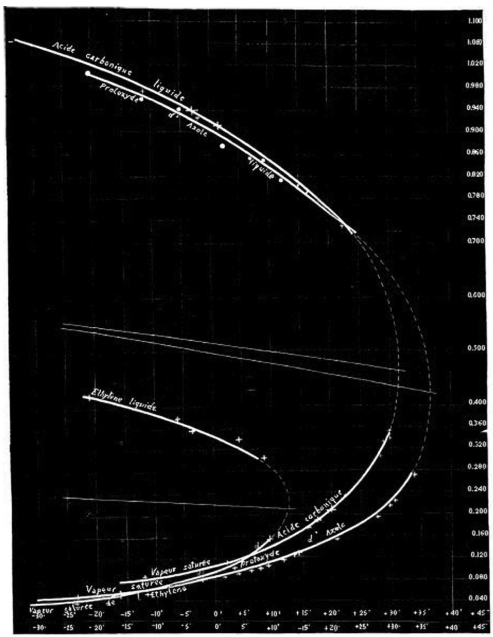} 
\caption{\label{lrd} 
Original data of the proposed law of rectilinear diameter by Cailletet and Mathias
for nitrous oxide, ethylene, and carbon dioxide 
(taken from reference \cite{cail-mat}). In the abscissa one has the temperature in the Celsius scale and in the ordinate the density in $10^{-3}$ kg/m$^3$. For each compound, the upper curve is the density of the liquid, and the lower curve the density of the vapor. }
\end{figure}
%
The critical density is then obtained from the measurement of $T_c$ as $\rho_c=\rho_o+A T_c$. The reader interested in the further developments and details of the empirical law of the rectilinear diameter is directed to the recent review by Reif-Acherman \cite{simon}. 

%
\begin{figure}[ht]
\includegraphics[clip,angle=0,width=6.5cm]{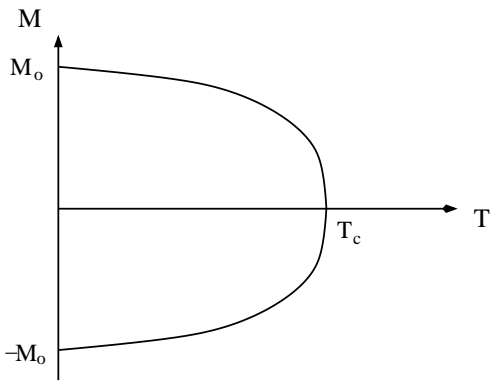} 
\caption{\label{mxt} 
Sketch of the magnetization $M$ as a function of temperature $T$, at zero external field, for a simple magnet. $M_o$ is the saturation magnetization and $T_c$ is the critical temperature.}
\end{figure}
%

Besides the study of the phase diagram of several substances, in which triple and critical points have been determined, the phase transitions of some magnetic materials were also well established by the end of the 19th century. In fact, in 1895, Pierre Curie (1859-1906) put the correspondence between magnets and fluids on a solid basis by taking the pressure $p$ as the analogue of the external magnetic field $H$, and the density $\rho$ as the analogue of the magnetization $M$ \cite{domb}. As a simple example, the sketch in Figure \ref{mxt} for the magnetization as a function of temperature for a simple symmetric magnet clearly shows the similarity to the density versus temperature for a fluid in Figure \ref{lrd}. By {\it symmetric} here one means that, at zero external field, the free energy is the same whether the magnetization  points in one direction or the other. This symmetry for the magnet, however, implies that the equivalent of the diameter for the magnetization is temperature independent, and any theory applied close to the critical point should be able to describe such diverse behavior. 

An astonishing discovery came quite later regarding the universal behavior of different substances close to a continuous phase transition. Measurable thermodynamic quantities, such as the order parameter $m$ (which can be taken as the magnetization $M$ for the magnet and the difference $\rho_L-\rho_V$ for the fluid), specific heat at constant volume $c_V$, magnetic susceptibility $\chi$ (or the compressibility in the fluid case), 
present a power law behavior of the form
\begin{eqnarray}
\nonumber
m\approx &m_o|t|^\beta,\\
\nonumber
c_V\approx &C_o|t|^{-\alpha},\\
\nonumber
\chi \approx &\chi_o|t|^{-\gamma},\\
\label{exp}
m\approx &m_1H^{1/\delta},
\end{eqnarray}
where ``$\approx$" means {\it asymptotically equal to}. In the above equations, $m_o$, $C_o$, $\chi_o$, and $m_1$ are constants, $\beta$, $\alpha$, $\gamma$, and $\delta$ are the corresponding thermal critical exponents, and $t=T-T_c$ (for questions of clarity, and without loss of generality, in what follows we will use this definition of $t$ instead of the more usual one $t=(T-T_c)/T_c$). More (spatial) exponents can be defined for the correlation function (exponent $\eta$), correlation length (exponent $\nu$), and also for the dynamical critical slowing down (dynamical exponent $z$). However, in the present work, for simplicity, we are going to treat only the thermal exponents in Eqs. (\ref{exp}). What happens is that, regardless of the microscopic structure of the interactions among the constituents of the compounds, the critical exponents are the same for systems belonging to the same universality class. A universality class, according to Kadanoff, can be defined primarily by the spatial  dimension of the lattice, the symmetry of the order parameter and the range of interactions \cite{stanley,kada}. One has also to consider the number of phases becoming equal at the continuous transition. In addition, the scaling laws also provide relations among critical exponents in such a way that knowing just two of them allows one to determine all the remainders. For example, just to cite a couple of them due to the works of Rushbrooke and Widom, we have\cite{stanb}
\begin{eqnarray}
\label{sre}
\alpha+2\beta+\gamma=2;&~~(\text {Rushbrooke})
\nonumber \\
\beta\delta=\beta+\gamma.&~~(\text {Widom})
\end{eqnarray}

A theoretical approach which could give a satisfactory picture of the scaling laws was proposed by Widom and some other workers (see, for instance, reference \cite{stanley} and references therein). The Widom scaling hypothesis just states that the singular part of the free energy is a generalized homogeneous function of its variables. From this hypothesis all the scaling laws follow, but the generalized homogeneous function does not provide the values of the critical exponents. The renormalization group (RG) approach of Wilson \cite{wilson,fisher} could give a complete microscopic account for the Widom hypothesis and the Kadanoff universality ideas, as well as turned out to be a quite powerful way of getting numerical values for the critical exponents. The RG can indeed be considered a cornerstone in the study of critical phenomena.

It should also be stressed that not only are the critical exponents universal but so are the scaling functions of the corresponding thermodynamic variables. For example, the universal character of the order-parameter probability distribution function (PDF) has been independently introduced by Bruce \cite{bruce} and Binder in 1981\cite{binder}. Its use in describing critical properties of models in statistical mechanics has been shown to be quite valuable and has also been extended in the study of pure and disordered magnetic systems, e.g. Lennard-Jones fluids, critical point in the    unified theory of weak and electromagnetic interactions and quantum chromodynamics, non-equilibrium phase transitions, among others. Due to the universal behavior of the order-parameter PDF, it becomes a fingerprint of the criticality of each class  of systems. If a match with this distribution is achieved, not only are the critical exponents in the same universality class, but also non-universal behavior (such as the determination of the corresponding critical temperature) can be readily extracted. 

Without a doubt, one of the best ways to compute the order-parameter PDF presented above is the use of Monte Carlo (MC) computer methods\cite{landa}. As was said previously in this Introduction, the application of the new instrument called {\it computer} has in some sense revolutionized not only Physics but the sciences in general. For more than half a century, computer simulations have been shown to be useful, and even crucial, to the understanding of most features and important subjects in modern science.  Indeed, in what concerns the present subject, MC simulations have been proven to be extremely valuable for obtaining information about phase transitions, regarding their universal and non-universal aspects, such as critical exponents, critical amplitudes, critical temperature, and even correction-to-scaling properties. The essential idea behind an MC procedure is to simulate a finite real system by generating a sample of states in which it can be found. However, for a very large system, only a small fraction of the total number of possible states is sampled. This leads, of course, to the inconvenience of having statistical errors associated with the physical quantities that are
evaluated during the simulation. Nevertheless, with the fast increase of computer power in recent decades and the development of more efficient algorithms, high-precision estimates can be achieved with sufficiently large runs. Furthermore, as computer simulations consider only finite systems, one must have a mechanism to extract the critical behavior (in the thermodynamic limit) from the size dependence of the measured physical quantities. In other words, it should be possible to get the non-analytical behavior of the infinite system at criticality by just studying the corresponding analytical behavior of finite systems. It turns out that such a procedure, in some cases, works quite well even for systems not too large. According to the finite-size scaling theory, proposed by Fisher in 1971, since near the transition the correlation length is limited by the system size $L$, one can get a set of relations similar to Eqs. (\ref{exp}) which includes the dependence on $L$ as well\cite{fss1,fss2}. For instance, at criticality ($T=T_c$) the extension of Eqs. (\ref{exp}) for the finite-size effects of the systems can be written as
\begin{eqnarray}
\nonumber
m\approx &m^\prime_oL^{-\beta/\nu},\\
\nonumber
c_V\approx &C^\prime_oL^{\alpha/\nu},\\
\label{fss}
\chi \approx &\chi^\prime_oL^{\gamma/\nu},
\end{eqnarray}
where $m^\prime_o$, $C^\prime_o$, and $\chi^\prime_o$ are constants. For any MC simulations, the finite-size scaling must be taken into account to avoid misleading results. Furthermore, one has also to spend some time in evaluating the errors coming from the simulations and the time correlations between configurations due to the pseudo-random number generators.  

We will herein discuss the usefulness of the probability distribution function of the order parameter, or any other conjugate extensive variable, in obtaining critical, and mainly multicritical, behavior of magnetic physical systems. Emphasis will be given on Monte Carlo simulations to compute the order-parameter distribution function, or any other needed distribution. We will discuss symmetric and (more ubiquitous) asymmetric cases. In order to get insight of the role of asymmetry in some models we will start with a brief revision of the mixing fields for fluid systems in the next section. We will closely follow the {\it revised scaling} by Rehr and Mermim \cite{rm} and the corresponding field mixing procedure introduced by Bruce and Wilding \cite{bw1,bw2}, since for the present systems it is not necessary to consider the general approach of the {\it complete scaling} proposed by Kim and Fisher in the study of highly asymmetric fluids \cite{kim1,kim2,kim3}.  The order parameter PDF in the symmetric case with its finite-size scaling relation will be derived in section III. In section IV we will apply it to the simplest Ising and Baxter-Wu models. A generalization of the finite-size scaling relation taking into account the presence of asymmetric fields will be discussed in section V. Examples of discrete spin systems as the Blume-Capel and spin-1 Baxter-Wu models, together with the Lennard-Jones fluid, will be discussed in section VI. In the last section, the use of the probability distribution to some other related models will be briefly presented together with some final remarks. 

\section{Free energy scaling function}
\label{mf}

\begin{figure}[ht]
\includegraphics[clip,angle=0,height=6.5cm]{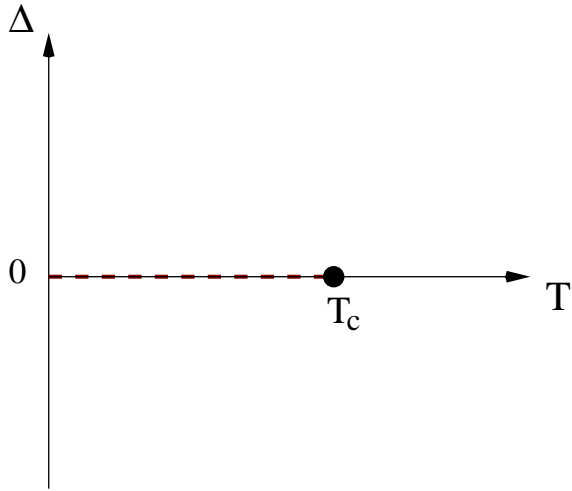} 
\caption{\label{sympd} 
Sketch of the symmetric phase diagram in the temperature field plane. The full circle is the critical point. First-order phase transitions take place for $T<T_c$ and $\Delta=0$. This is the equivalent field phase diagram of Figure \ref{mxt}.}
\end{figure}

For any symmetric system it is well known that the free energy $g(T,\Delta)$, in terms of the temperature $T$ and the {\it field} $\Delta$, can be written as 
\begin{equation}
\label{sg}
g(T,\Delta)= g_o(T,\Delta)+|t|^{2-\alpha}f_\pm(\Delta/|t|^{\beta\delta}),
\end{equation}
where $g_o(T,\Delta)$ is an analytic function, $t=(T-T_c)$ and the scaling function $f_+$ holds for $T>T_c$ and $f_-$ for $T<T_c$, with the exponents defined in Eqs. (\ref{exp}). The symmetry here means that the critical point is located at $t=0$ and $\Delta=0$. This is the case for the Ising model where we have the correspondence $\Delta=H$, $H$ being the external magnetic field, as depicted in Figure \ref{sympd}.
%
%
%

\begin{figure}[ht]
\includegraphics[clip,angle=0,width=8.2cm]{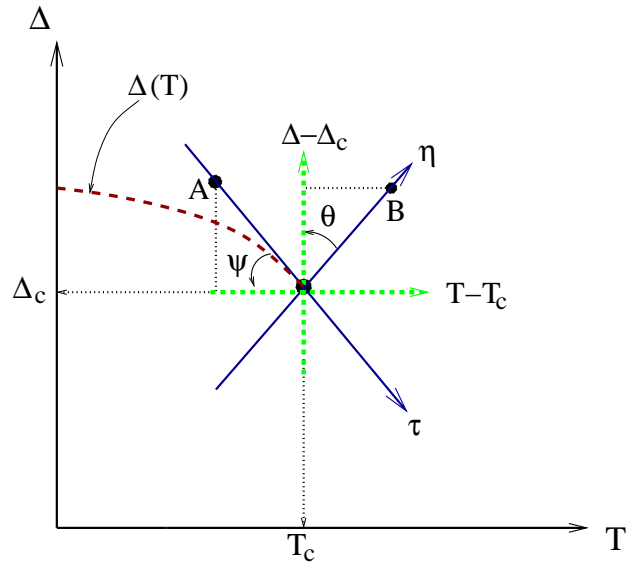} 
\caption{\label{DxT} 
Sketch of the first-order transition line $\Delta(T)$ in the field plane. ($T_c,\Delta_c$) is a critical or, in general, a multicritical point when one has an additional second-order transition line. $\tau$ and $\eta$
are the field mixing directions. All points ($T,\Delta$) along the $\tau$ direction
(such as point $A$) have $\eta=0$ while points like $B$ have $\tau=0$.
}
\end{figure}

However, it is more common to have a first-order transition line where the field $\Delta$ is a function of the temperature $T$, namely  $\Delta=\Delta(T)$. This is the case, for instance, in the Blume-Capel model where $\Delta$ plays the role of the crystal field (or single ion anisotropy), and in fluid systems where $\Delta$ can be the pressure or chemical potential. A typical phase diagram is sketched in Figure \ref{DxT} and a possible generalization of the scaling function (\ref{sg}) can be written as
\begin{equation}
\label{asg}
g(T,\Delta)=g_o(T,\Delta)+|t|^{2-\alpha}f_\pm[(\Delta-\Delta(t))/|t|^{\beta\delta}].
\end{equation}
%
%
%

The above scaling function, as well as its extension by including mixing fields, has been extensively studied by Rehr and Mermin \cite{rm} in the context of the fluid critical point.  In reference \cite{rm}, instead of the free energy, the pressure, as a function of temperature and chemical potential, has been considered. As the Gibbs free energy per particle of a fluid is the chemical potential $\mu$ written in terms of the intensive temperature and pressure variables, i.e. $\mu=\mu(T,p)$, this equation can be inverted in order to get $p=p(T,\mu)$. In doing this one gets equations of state of the form ${{s}\over{v}}=\left({{\partial p}\over{\partial T}}\right)_\mu$ and $\rho={{1}\over{v}}=\left({{\partial p}\over{\partial \mu}}\right)_T$, where the entropy per particle is $s=S/N$, $v=V/N$ is the volume per particle, and $\rho$ is the (number) density. In addition, the isothermal compressibility $\kappa_T=-{{1}\over{V}}\left({{\partial V}\over{\partial p}}\right)_T$ can also be given by $\kappa_T={{1}\over{\rho^2}}\left({{\partial \rho}\over{\partial \mu}}\right)_T$. This implies that the scaling function (\ref{asg}) should be compatible with the power law behavior of the order parameter and compressibility given in Eqs. (\ref{exp}) as well as with the rectilinear diameter law in (\ref{lrd}). 

Let us for the moment consider $g\equiv p$ and $\Delta\equiv \mu$ in order to treat in more detail the fluid system according to (\ref{asg}). As discussed in reference \cite{rm}, for $T<T_c$ the function $f_-(z)$ is an even function of the variable $z$ with the liquid phase holding for $z\rightarrow 0^+$ and the vapor phase for $z\rightarrow 0^-$. We then have
\begin{equation}
\label{rho}
\rho=\left({{\partial g}\over{\partial \Delta}}\right)_t=\rho_c+|t|^\beta f_-^\prime(z),
\end{equation}
\begin{equation}
\label{kap}
\rho_c^2\kappa_t=\left({{\partial \rho}\over{\partial \Delta}}\right)_t=|t|^{-\gamma} f_-^{\prime\prime}(z),
\end{equation}
where we have used Eqs. (\ref{sre}) in order to get $2-\alpha-\beta\delta=\beta$ and $2-\alpha-2\beta\delta=-\gamma$. It is clear from Eq. (\ref{kap}) that the compressibility indeed satisfies the corresponding power law behavior expressed in Eqs. (\ref{exp}). For the density Eq. (\ref{rho}) gives
\begin{equation}
\label{rholv}
\rho_L=\rho_c+|t|^\beta f^\prime_-(0^+),~~\rho_V=\rho_c+|t|^\beta 
       f^\prime_-(0^-),
\end{equation}
and as $f^\prime_-(0^+)=-f^\prime_-(0^-)$ one has
\begin{equation}
\label{op}
\rho_L-\rho_V=2f^\prime_-(0^+)|t|^\beta,
\end{equation}
meaning that the fluid order parameter $\rho_L-\rho_V$ also possesses its expected scaling behavior. However, from Eqs. (\ref{rholv}) one cannot obtain the law of rectilinear diameter in (\ref{lrd}) unless $A=0$. This is alright for the symmetric case depicted in Figure \ref{mxt} where the corresponding diameter of the magnetization is temperature independent. Although corrections to scaling in Eq. (\ref{asg}) can provide the expected behavior for the diameter in asymmetric systems, Rehr and Mermin \cite{rm} have proposed a field mixing procedure which is not so drastic as the corrections to scaling. Instead of working with the natural fields $T-T_c$ and $\Delta-\Delta_c$, as illustrated by the straight dashed lines in Figure \ref{DxT}, one can define more appropriate fields in the following way. 

Consider the first-order transition line in the $T\times \Delta$ plane in such a way that it is given by a function $\Delta(T)$. Close to the multicritical point one has
\begin{equation}
\Delta(T)=\Delta_c-r(T-T_c)+...,
\label{DT}
\end{equation}
where $r=\tan\Psi$ (see Figure \ref{DxT}). If we neglect the higher-order
terms in Eq.(\ref{DT}) we have
\begin{equation}
(\Delta-\Delta_c)+r(T-T_c)= 0.
\label{DT0}
\end{equation}
The equation above gives a straight line tangent to the $\Delta(T)$ curve at ($T_c,\Delta_c$), so we can define a field
\begin{equation}
\eta=(\Delta-\Delta_c)+r(T-T_c),
\label{eta}
\end{equation}
which may have any direction not parallel to the tangent line. Similarly, we have
\begin{equation}
\tau=(T-T_c)-s(\Delta-\Delta_c),
\label{tau}
\end{equation}
where $s=\tan\theta$. $\eta=0$ defines the $\tau$ axis while $\tau=0$ defines the
$\eta$ axis, as shown in Figure \ref{DxT}. When $r=s=0$ one gets the symmetric
choice ($\tau=T-T_c,\eta=\Delta-\Delta_c$). Thus, in these new fields Eq. (\ref{asg}) takes the form
\begin{equation}
\label{asgmf}
g(t,\Delta)=g_o(\tau,\eta)+|\tau|^{2-\alpha}f_\pm(\eta/|\tau|^{\beta\delta}).
\end{equation}
It should be noted that in the language of a fluid $\eta\rightarrow0^+$ means the liquid phase while $\eta\rightarrow0^-$ means the vapor phase. From Eq. (\ref{asgmf}) one gets
\begin{equation}
\label{opmf}
\rho_L-\rho_V=2{{\partial \eta}\over{\partial \mu}}f^\prime_-(0^+)|t|^\beta,
\end{equation}
%
%
%
%
\begin{equation}
\label{rhod}
\rho_d=\rho_c+(2-\alpha){{\partial |\tau|}\over{\partial \mu}}f_-(0^+)|\tau|^{1-\alpha}.
\end{equation}
This means that the scaling relation (\ref{asgmf}) indeed yields the singularities observed in the order parameter and in the diameter. Regarding the diameter itself one sees that the temperature exponent is not $1$. As $\alpha$ is small the difference is not so large. Nevertheless, some experiments do show deviations from Eq. (\ref{lrd}) and quite a few different empirical approaches have been proposed to deal with them (the interested reader should consult reference \cite{cail-mat}). Moreover, the singularities of the isothermal compressibility, specific heat, as well as the difference $\left({{\partial \rho}\over{\partial \mu}}\right)_L-\left({{\partial \rho}\over{\partial \mu}}\right)_V\propto |t|^{\beta-1}$, which are omitted here for simplicity, are also reproduced by the scaling relation with the mixed fields in Eq. (\ref{asgmf}). It should be said that the latter difference is {\it not} obtained from the simple relation (\ref{asg}) either \cite{rm}. Additional contributions to Eq. (\ref{rhod}) are obtained when one considers the complete scaling procedure\cite{text}. 

Thus, whenever one deals with models having asymmetries in the phase diagram, one has to work with the mixing fields given by Eqs. (\ref{eta}) and (\ref{tau}) rather than $(t,\Delta)$. This procedure can also be understood as a way of minimizing the corrections to scaling, always present in experiments for not being close enough to the critical or multicritical point, as well as in simulations for not being able to simulate large enough lattices. 

In what follows we will discuss these ideas with the use of the order-parameter probability distribution function in getting not only the universality class of the system but also locate the multicritical point, which is really a difficult task in some cases.

\section{Order parameter probability distribution function} 

For an infinite system undergoing a continuous phase transition at a temperature $T_c$ one has the non-analytic behavior expressed by Eqs. (\ref{exp}) for those thermodynamic quantities. On the other hand, for a finite system of linear dimension $L$, with a number of sites $N=L^d$, where $d$ is the dimension of the lattice, those singularities are smeared out and an analytic behavior is observed instead. For rather small values of $L$ exact results (partition function) can be obtained. However, as $L$ increases the number of states of the system gets drastically larger in such a way that exact results become impractical, in some cases, even for system sizes of around a dozen particles. In this way, Monte Carlo (MC) simulations turn out to be one
of the most efficient methods to measure the thermodynamical variables like the magnetization or order parameter $m$ of a magnetic system, and also its probability distribution $P_L(m,T)$. 

At the critical temperature and for very large system sizes ($L>>1$) one expects the 
following scaling relation \cite{bruce,binder}
\begin{equation}
P_L(m) =bL^{\beta/\nu}P^*(bL^{\beta/\nu}m),
\label{ssr}
\end{equation}
where $\nu$ is the correlation length critical exponent, $b$ is a non-universal metric constant, and $P^*$ is a universal function.  $P^*$ is a fingerprint of the corresponding  universality class. Note that in the above equation $m$ is considered a continuous variable. The task of computing $P^*$ from Monte Carlo simulations and, in addition, its utility in obtaining either the criticality or the multicriticality of symmetric and asymmetric models, will be exemplified in the next sections.

\section{Probability distribution in the symmetric case: Ising and Baxter-Wu models}

Let us start with the symmetric Ising model which can be defined by the following Hamiltonian
\begin{equation}
{\cal H} = -J\sum_{\langle i,j\rangle} S_i S_j-H\sum_{i=1}^N S_i,
\label{ih}
\end{equation}
where the first sum is over nearest-neighbor pairs of spins $\langle i,j\rangle$ and the last
one is over all $N=L^d$ sites of a hyper cubic $d$-dimensional lattice with linear size $L$. $J$ is the  exchange interaction and $H$ is the external magnetic field. The spins $S_i$ can take values 
$-S, -S+1,...,S-1,S$. The simplest case is the spin-1/2 model where $S=1/2$.

For a finite-size system of linear dimension $L$ and a Monte Carlo run with ${\cal N}$ 
Monte Carlo steps per spin (after discarding some initial 
configurations for equilibrating the system) one has, for each generated configuration $k$, 
a measure of the energy at zero field $E_L^k$ and the corresponding magnetization $M_L^k$
\begin{equation}
E_L^k=\sum_{\langle i,j\rangle} S_i S_j,~~~~~~~M_L^k= \sum_{i=1}^N S_i,
\label{emih}
\end{equation}
in such a way that the Ising Hamiltonian (\ref{ih}) can be written as
${\cal H}_L=-JE_L^k-HM_L^k$. One then has the following mean values for 
$E_L$, $M_L$ and ${\cal H}_L$
\begin{eqnarray}
\left<E_L\right>&=&{{1}\over{{\cal N}}}\sum_{k=1}^{{\cal N}} E_L^k,~~~~
\left<M_L\right>={{1}\over{{\cal N}}}\sum_{k=1}^{{\cal N}} M_L^k,
\nonumber\\
\left<{\cal H}_L\right>&=&-J\left<E_L\right>-H\left<M_L\right>.
\label{memh}
\end{eqnarray}

For this run one can also compute $P_L(M_i)$, which is the probability of having the
magnetization $M_i$ in this particular simulation (in this case $-SN\le M_i\le SN$)
\begin{equation}
P_L(M_i)={{number~of~occurrences~of~M_i}\over{{\cal N}}}.
\label{pm}
\end{equation}
Note that the mean value of $M_L$ is also given by
\begin{equation}
\left<M_L\right>=\sum_{i} M_iP_L(M_i),
\label{mdm}
\end{equation}
as expected, and from Eq.(\ref{pm}) one also has the normalization condition
\begin{equation}
\sum_iP_L(M_i)=1.
\label{n1}
\end{equation}
In the more general case one can obtain the joint probability distribution
$P_L(E_i,M_i)$. The distribution given by Eq.(\ref{pm}) is obtained from
$P_L(M_i)=\sum_{E_i}P_L(E_i,M_i)$,
where $-S^2Nc/2\le E_i\le S^2Nc/2$, with $c$ the coordination number of the lattice.

The next step is now to compute $P_L(m)$ from the measured $P_L(M_i)$.
Eq.(\ref{n1}) can be rewritten as
\begin{equation}
\sum_i{{P_L(M_i)}\over{\Delta M}}\Delta M=1,
\label{n2}
\end{equation}
where $\Delta M=1$ is the discrete magnetization step
(if, as it is usual for the spin-1/2 Ising model, one defines $S_i=\pm 1$ one would have
$\Delta M=2$). By defining $m_i={{M_i}\over{SN}}$, one has a discrete series of non-integer 
values $-1\le m_i\le 1$, so the above equation can be written as
\begin{equation}
\sum_{m_i=-1}^{m_i=1}{{P_L(m_i)}\over{\Delta M/SN}}{{\Delta M}\over{SN}}=
\sum_{m_i=-1}^{m_i=1}{{P_L(m_i)}\over{\Delta M/SN}}dm_i=1.
\label{n3}
\end{equation}
Now, for a large system the quantity $dm_i={{\Delta M}\over{SN}}$ is very small 
and the magnetization $m_i$ can be regarded as a continuous variable so the
above sum can be replaced by the integral
\begin{equation}
\int_{-1}^{1}P_L(m)dm=1,
\label{n4}
\end{equation}
where
\begin{equation}
P_L(m)={{P_L(m_i)}\over{\Delta M/SN}},
\label{n5}
\end{equation}
which is the desired expression for the continuous probability distribution.

This probability distribution satisfies the scaling relation (\ref{ssr})
\begin{equation}
P_L(m)=bL^{\beta/\nu}P_L^*(bL^{\beta/\nu}m)=bL^{\beta/\nu}P_L^*(m^*),
\label{psr}
\end{equation}
where $m^*=bL^{\beta/\nu}m$.

The normalization condition
\begin{equation}
\int P_L(m)dm=1=\int bL^{\beta/\nu}P_L^*(m^*)dm=\int P_L^*(m^*)dm^* 
\end{equation}
implies that $P_L^*(m^*)$ is already normalized.

On the other hand, the variance is given by
\begin{equation}
\sigma^2=\int m^2P_L(m)dm,
\end{equation}
\begin{equation}
\sigma^2=\int m^2P_L^*(m^*)dm^*={{1}\over{b^2L^{2\beta/\nu}}}\int {m^*}^2P_L^*(m^*)dm^*,
\end{equation}
\begin{equation}
\sigma^2={{{\sigma^*}^2}\over{b^2L^{2\beta/\nu}}}.
\end{equation}
Thus, also normalizing the variance ${\sigma^*}^2=1$, one has $bL^{\beta/\nu}=1/\sigma$
and $m^*=m/\sigma$ and one gets rid of the scaling exponent 
\begin{equation}
P_L^*(m/\sigma)=\sigma P_L(m),
\label{psrn}
\end{equation}
where $m$, $P_L(m)$ and $\sigma$ are measured in an MC simulation. In this way we obtain the universal normalized function $P^*$ with unit variance.

\subsection{One-dimensional Ising Model}

Exact results can be obtained for the one-dimensional Ising model \cite{bruce}. On the other hand, analytic results can be obtained as well in higher dimensions by using some approximate approach \cite{bruce}. In the present case, however, we will be mostly concerned with the order-parameter PDF computed from Monte Carlo simulations. Such a procedure has been shown to be very useful not only for the PDF itself but also in order to get the transition point. Details for the exact PDF for the one-dimensional model using the transfer matrix technique can be found in reference \cite{bruce}.

\subsection{Two-dimensional Ising Model}

\begin{figure}[ht]
\includegraphics[clip,angle=-90,width=9.5cm]{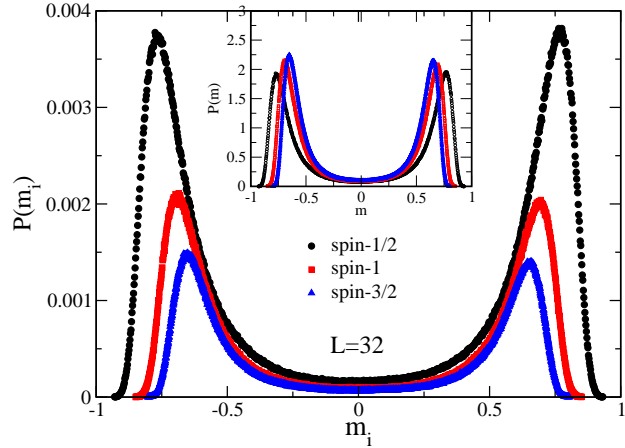} 
\caption{  Probability distribution function $P(m_i)$, where $m_i=M_i/NS$, according to Eq. (\ref{pm}) for the two-dimensional Ising model with spin-$1/2,1$, and $3/2$. The data correspond to lattice size $L=32$. The exact $T_c=2.2692...$ has been taken for the spin-$1/2$, $T_c=1.6935$ for the spin-$1$ \cite{adler} and $T_c=3.2879$ for the spin-$3/2$ \cite{pla1}. The inset shows the corresponding distribution $P(m)$ in the continuum limit from Eq. (\ref{n5}) for the same lattice and spin values. The error bars are smaller than the symbol sizes.}
\label{PMim123}
\end{figure}

\begin{figure}[ht]
\includegraphics[clip,angle=0,width=8.3cm]{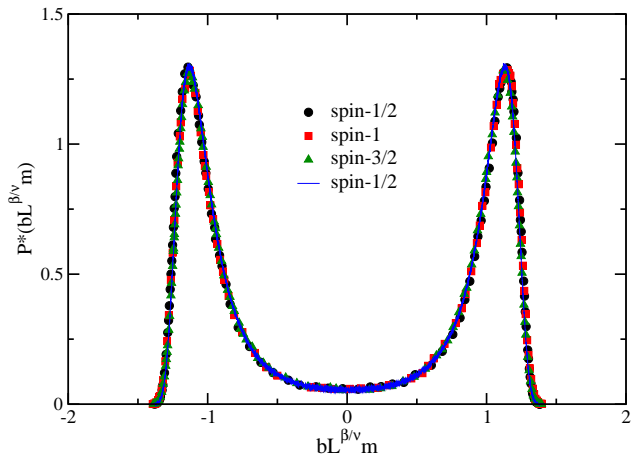} 
\caption{  Universal $P^*(m^*)$ distribution function as function of $m^*=bL^{\beta/\nu}m$ for the two-dimensional Ising model with spin-$1/2,1$, and $3/2$ according to Eq. (\ref{psrn}). In order to have a clearer picture, lesser data than presented in Figure \ref{PMim123} has been taken in this case. The extra solid line corresponds to a Wolff algorithm to the spin-$1/2$ model for a lattice size $L=100$ with $10^8$ MCS. The error bars are smaller than the symbol sizes.}
\label{is123}
\end{figure}

The two-dimensional universality class of the PDF for the Ising model has been studied by Binder \cite{binder} and Nicolaides and Bruce \cite{nico}. As a matter of illustration, Figure \ref{PMim123} shows the probability distribution of $P_L(m_i)$ according to Eq. (\ref{pm}), where the order parameter has been normalized by $m_i=M_i/NS$. In this simple example one has a rather small latice $L=32$. The method used was the Metropolis algorithm for spin-$1/2$, spin-$1$, and spin-$3/2$ with just $6\times10^6$ MCS per spin. The inset shows the corresponding continuous limit of the probability distribution resulting from taking Eq. (\ref{n5}). Different curves are obtained for different values of $S$. Finally, Figure \ref{is123} depicts the two-dimensional Ising probability distribution function of the order parameter given by Eq. (\ref{psrn}), where one can clearly see the collapse 
%
%
of all the curves.  For comparison, in Figure \ref{is123} one also has  the data from the Wolff algorithm applied to the spin-$1/2$ Ising model for a lattice size $L=100$ with $10^8$ MCS. One can notice that, in this case, there is a very definite universal function which is independent of the spin value, as expected. Larger lattices and longer MCS will provide quite similar results with smaller finite-size corrections and smoother curves.
%
%

Figure \ref{is123} clearly shows the universal character of the probability distribution function $P^*$, which could be computed by knowing, {\it a priori}, the transition temperature. In spite of that a different approach has been proposed to get not only the critical temperature but also the order-parameter PDF in cases where one does not know the exact location of the continuous transition \cite{puli1}. One first generates the PDF $P_{L_o}(m,T_o)$, where $T_o$, although close to the actual transition temperature, can be either above or below $T_c$. Simulations are then performed to obtain a matching of $P_L(m,T_L)$ with $P_{L_o}(m,T_o)$ for $L\ne L_o$ at a temperature $T_L\ne T_o$. A usual finite-size scaling analysis is then used to get $T_c$ and the exponent $\nu$, which in turn allows one to get the desired universal function $P^*$. This has been applied to the two-dimensional spin-$1/2$ and spin-$1$ Ising models \cite{puli1}.

Knowledge of $P^*$ for this universality class can also be very useful in treating the diluted two-dimensional Ising model. In this case, the Harris criterion is not fulfiled and there is not yet a consensus whether or not the critical exponents change with dilution. By employing MC simulations and using $P^*$ as a guide, it was possible to analise the corresponding order-parameter PDF \cite{puli2}. The results agree with the strong universality scenario, which predicts only logarithmic corrections to the critical behavior with the same critical exponents as the pure system. Regarding this open question, more recent simulations also agree with the exponents being the same as the pure model \cite{tasos}.
 
\subsection{Three-dimensional Ising Model}

\begin{figure}[ht]
\includegraphics[clip,angle=0,width=8.3cm]{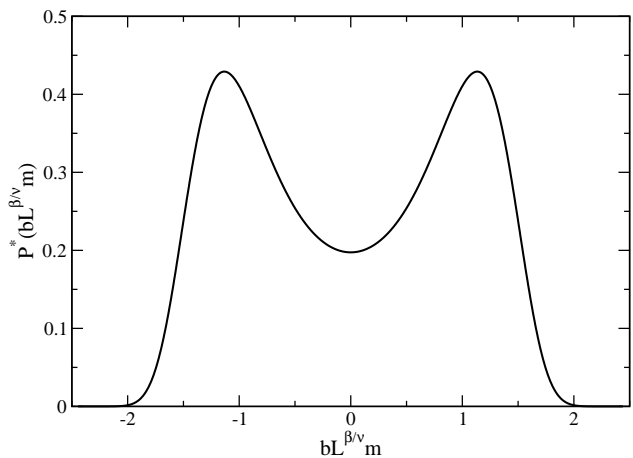} 
\caption{  Analytical $P^*$ distribution function for the three-dimensional Ising model according to reference \cite{tsypin} using the results for lattice $L=20$.}
\label{is3d}
\end{figure}

Figure \ref{is3d} shows the order-parameter PDF for the three-dimensional Ising model universality class according to reference \cite{tsypin}. In spite of some progress in obtaining this function by analytical methods this is a nice example where simulations are the main source of information about the PDF properties. The curve in Figure \ref{is3d} is given by Eq. (3) of reference \cite{tsypin} using the fitting data for the  lattice $20^3$.
%
%

\subsection{Spin-$1/2$ Baxter-Wu Model}

\begin{figure}[ht]
\includegraphics[clip,angle=0,width=8.3cm]{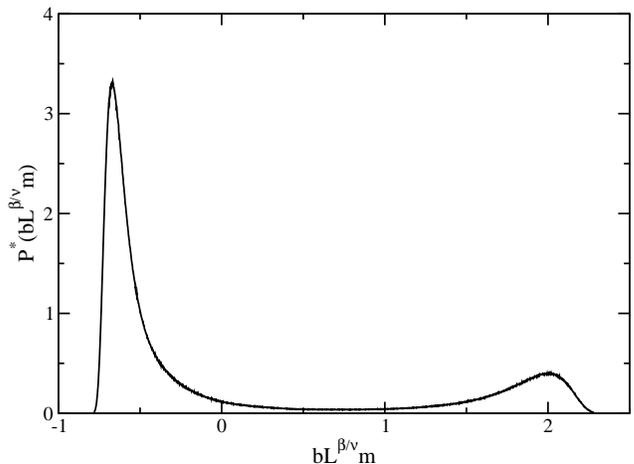} 
\caption{  $P^*$ distribution function for the two-dimensional spin-$1/2$ Baxter-Wu model. The noise in the data give an idea of the error in the simulation.}
\label{bws05}
\end{figure}

The Baxter-Wu model can be defined by the following Hamiltonian
\begin{equation}
{\cal H} = -J\sum_{i,j,k} S_i S_jS_k,
\label{bw}
\end{equation}
where the sum is over all triangles on a triangular lattice. The exact solution given by Baxter and Wu \cite{bw} shows a continuous transition at the same critical temperature as the two-dimensional Ising model with critical exponents belonging to the 4-state Potts model universality class.
The corresponding order-parameter PDF for the spin-$1/2$ Baxter-Wu model (when the variables are $S_i=\pm1$) is depicted in Figure \ref{bws05}. The data in Figure \ref{bws05} have been obtained for a lattice size $L=60$ at the exact critical temperature $k_BT_c/J=2.2692$ using Metropolis algorithm with 25 million MCS. This distribution is similar to that obtained in reference \cite{tasos2} (the distribution shown in Figure \ref{bws05} has its variance normalized according to the present text). Since the Baxter-Wu model corresponds to three-spin interactions on a triangular lattice, the ground state has one ferromagnetic phase with magnetization $m=1$ and three ferrimagnetic phases with magnetization $m=-1/3$. This is the reason of having one peak at $m$ and another more pronounced peak at $-m/3$. 
%
%

\section{Extensive Mixing Fields Variables}

So far we have considered systems presenting symmetric phase diagrams. In such cases, as for the Ising model described by the Hamiltonian (\ref{ih}) one has the partition function
\begin{equation}
Z=\sum_{states~\ell}e^{\beta J\sum_{\langle i,j\rangle} S_i S_j
+\beta H\sum_{i=1}^N S_i}=\sum_{states~\ell}e^{\tilde{t}E_\ell+\tilde{h}M_\ell},
\label{pfih}
\end{equation}
where $E_\ell=\sum_{\langle i,j\rangle} S_i S_j$ and $M_\ell=\sum_{i=1}^N S_i$ are the energy and magnetization of the labeled state $\ell$. We have also introduced the variables $\tilde{t}=\beta J$ and $\tilde{h}=\beta H$, which are the inverse reduced temperature and reduced
external field, respectively, where $\beta=1/k_BT$ with $k_B$ the Boltzmann constant (this definition of $\beta$ should not be confused with the order parameter critical exponent previously defined). In addition, from above one has the mean values
\begin{equation}
\left<E\right>={{1}\over{Z}}{{\partial Z}\over{\partial \tilde t}},~~~~~~~~~
\left<M\right>={{1}\over{Z}}{{\partial Z}\over{\partial \tilde h}},
\label{EMih}
\end{equation}
where the extensive quantities $E$ and $M$ are related to the {\it intensive fields}
$\tilde t$ and $\tilde h$, respectively.
The corresponding free energy is given by 
$G(T,H)=G(\tilde t,\tilde h)=-k_BT\text{ln}Z$. The second-order phase transition is located at $T_c$ and
$H=0$, so for this symmetric problem one has the {\it scaling fields} ($T-T_c$,~$H$) as in Figure \ref{sympd}. 

In most cases, however, the phase transition line in the field space is not
symmetric as in the Ising model. For instance, consider the Blume-Capel model defined by
\begin{equation}
{\cal H} = -J\sum_{\langle i,j\rangle} S_i S_j-H\sum_{i=1}^N S_i-\Delta\sum_{i=1}^N S_i^2,
\label{bch}
\end{equation}
with $\Delta$ being the crystal field and $S\ge 1$. In this case there is a first-order phase
transition line in the temperature versus crystal field plane (for $H=0$) ending at a multicritical point \cite{plazc}. The corresponding phase diagram regarding the first-order transition looks like that sketched in Figure \ref{DxT}. 
The corresponding partition function is given by
\begin{eqnarray}
Z&=&\sum_{states~\ell}e^{\beta J\sum_{\langle i,j\rangle} S_i S_j
+\beta H\sum_{i=1}^N S_i+\beta \Delta\sum_{i=1}^NS_i^2}
\nonumber\\
&=&\sum_{states~\ell}e^{\tilde{t}E_\ell+\tilde{h}M_\ell+\tilde qQ_\ell}.
\label{pfbch}
\end{eqnarray}
So, in addition to Eqs.(\ref{EMih}) we have here
\begin{equation}
\left<Q\right>={{1}\over{Z}}{{\partial Z}\over{\partial \tilde q}},
\label{mQ}
\end{equation}
where $\tilde q=\beta \Delta$ and $\left<Q\right>$ is the mean value of the square of the spins, or quadrupole moment, obtained from
\begin{equation}
Q_\ell=\sum_{i=1}^N S_i^2.
\label{Q}
\end{equation}

Close to the multicritical point (located on the plane defined by $H=0$) one would take as a possible and naive
choice the fields ($T-T_c,~\Delta-\Delta_c$) where,
for simplicity, the multicritical coordinates are designated by ($T_c,~\Delta_c,~H_c$). This choice, 
as has been discussed in section \ref{mf}, is not convenient and mixing fields have been proven to be more desirable to describe the scaling behavior close to the continuous transition. 

\subsection{General Approach}

Let us first discuss the general approach to this problem which is not applicable only to the Hamiltonian (\ref{bch}) but to all systems where the first order line is not located at constant field $\Delta$ (in this case $Q$ is the corresponding extensive quantity of the field $\tilde q=\beta\Delta$). Eqs. (\ref{eta}) and (\ref{tau}) can then be written in a more general form by  defining $\hat t=\tilde t - \tilde t_c$ and $\hat w=\tilde q-\tilde q_c$, so that one has
\begin{equation}
\eta=\hat w+r\hat t,
\label{etab}
\end{equation}
\begin{equation}
\tau=\hat t-s\hat w.
\label{taub}
\end{equation}
Note that, in this case, close to the multicritical point $(T_c,\Delta_c)$ instead of the axes $T-T_c$ and $\Delta-\Delta_c$ as in Figure \ref{DxT}, we have now $\hat t={{J}\over{k_BT_c}}{{T_c-T}\over{T_c}}$ and $\hat \omega={{\Delta-\Delta_c}\over{k_BT_c}}$, which differ by a metric factor from the former ones depicted in Figure \ref{DxT}.

The question now is how to derive the extensive quantities associated to the fields
$\eta$ and $\tau$ which are not explicitly present in the Hamiltonian (\ref{bch}).
Nevertheless, analogously to Eq.(\ref{EMih}) one can define
\begin{equation}
\left<\mathbb E\right>={{1}\over{Z}}{{\partial Z}\over{\partial \tau}},~~~~~~~~~
\left<\mathbb Q\right>={{1}\over{Z}}{{\partial Z}\over{\partial \eta}},
\label{EQ}
\end{equation}
and we get
\begin{eqnarray}
\left<\mathbb E\right>={{1}\over{Z}}\left({{\partial Z}\over{\partial \hat t}}
                    {{\partial \hat t}\over{\partial \tau}}+{{\partial Z}\over{\partial \hat \omega}}
                    {{\partial \hat \omega}\over{\partial \tau}}
\right), \nonumber\\
\left<\mathbb Q\right>={{1}\over{Z}}\left({{\partial Z}\over{\partial \hat t}}
                    {{\partial \hat t}\over{\partial \eta}}+{{\partial Z}\over{\partial \hat \omega}}
                    {{\partial \hat \omega}\over{\partial \eta}}
\right).
\label{part}
\end{eqnarray}
Inverting Eqs. (\ref{etab}) and (\ref{taub}) we obtain
\begin{equation}
\hat \omega ={{1}\over{1+rs}}(\eta-r\tau),
\label{d}
\end{equation}
\begin{equation}
\hat t={{1}\over{1+rs}}(\tau+s\eta).
\label{t}
\end{equation}
As the partial derivatives of the partition function with respect to $\hat t$ and $\hat \omega$ are identical to the ones with respect to $\tilde t$ and $\tilde q$, we can use the results (\ref{EMih}) and  (\ref{mQ}) to write the extensive variables in Eqs.(\ref{part}) as
\begin{equation}
{\mathbb E}={{1}\over{1+rs}}(E-rQ),
\label{E}
\end{equation}
\begin{equation}
{\mathbb Q}={{1}\over{1+rs}}(Q+sE).
\label{Q}
\end{equation}

The main task now is to get the probability distribution function of these extensive variables. As before, from an MC simulation on a finite lattice of linear size $L$, one is able to straightforwardly measure the joint probability distribution $P_L(E_i,Q_i)$ of having discrete values of energy $E_i$ and quadrupole moment $Q_i$. From this probability distribution we can obtain the corresponding probability distribution ${P}_L({\mathbb E}_i,{\mathbb Q}_i)$ for the discrete extensive operators ${\mathbb E}_i$ and ${\mathbb Q}_i$ for this lattice size. In fact, what we need for this problem is to get the probability distribution ${\cal P}_L({\cal Q})$ of the continuous variable ${\cal Q}$ obtained from the probability distribution of the discrete variable ${\mathbb Q}$. The procedure is quite similar to that described in the previous section for the magnetization of symmetric models.

In order to do that we first consider a discrete variable $x_i$ with normalized probability $P(x_i)$ where
\begin{equation}
\left<x\right>=\sum_ix_iP(x_i).
\label{d1}
\end{equation}
We also have
\begin{equation}
\left<ax\right>=\sum_iax_iP(x_i)=a\left<x\right>.
\label{d2}
\end{equation}
For the variable $y_i=ax_i$ one has a probability distribution $P^\prime(y_i)$ in such
a way that
\begin{equation}
\left<y\right>=\sum_iax_iP^\prime(ax_i)=a\left<x\right>,
\label{d3}
\end{equation}
which, comparing Eqs. (\ref{d2}) and (\ref{d3}),  gives $P^\prime(ax_i)=P(x_i)$. This means that one is able to get the probability distribution $P_L({\mathbb Q}_i)$ for the variable ${\mathbb Q}_i$ defined in Eq. (\ref{Q}) and, as we will see below, this is the key distribution for getting the desired universal one. As the MC data are stored in pairs of configurations $(E_i,Q_i)$, from these configurations we can easily construct the probability for the composed variables $Q_i+sE_i$ by just counting their corresponding number of occurrences in the total run. This probability is equal to the probability of the variable ${\mathbb Q}_i$. In this way the probabilities $P_L({\mathbb E}_i)$ and $P_L({\mathbb E}_i,{\mathbb Q}_i)$ could also be computed, if necessary. Note still that from Eq. (\ref{Q}) the variable ${\mathbb Q}_i$ is discrete, but not integer, and not equally spaced, in such a way that we cannot use something similar to Eq. (\ref{n5}) to get its continuous distribution. 

On the other hand, we can define the variable
\begin{equation}
{\cal Q}_i={{{\mathbb Q}_i}\over{{\mathbb Y}_{max}}}={{1}\over{1+rs}}{{(Q_i+sE_i)}\over{{\mathbb Y}_{max}}}={{1}\over{1+rs}}y_i,
\label{q}
\end{equation}
where ${\mathbb Y}_{max}$ is the maximum value of the variable ${\mathbb Y}_i=Q_i+sE_i$ and $y_i={{(Q_i+sE_i)}\over{{\mathbb Y}_{max}}}$. Note further that 
\begin{equation}
P_L({\cal Q}_i)=P_L(y_i)=P_L({\mathbb Q}_i),
\label{Pq}
\end{equation}
and $y_i$ can be viewed as a continuous variable ranging in the interval $-1< {y_i}\le 1$ (in general, the mean value of ${\mathbb Y}_i$ is not zero, and we are assuming the distribution is dislocated to the right --- a different value of ${\mathbb Y}_{max}$ can be taken when the distribution is dislocated to the left, in order to limit the $y_i$ variable to the interval $-1$ to $1$).

Now, for a continuous distribution $P(x)$ one has
\begin{equation}
\int_{-{\infty}}^{\infty}P(x)dx=1,~~\int_{-{\infty}}^{\infty}xP(x)dx=\left<x\right>.
\label{}
\end{equation}
For the probability distribution ${\cal P}(y)$ with $y=ax$ one has
\begin{equation}
\int_{-{\infty}}^{\infty}{\cal P}(y)dy=\int_{-{\infty}}^{\infty}{\cal P}(ax)adx=1,
\label{}
\end{equation}
which gives
\begin{equation}
a{\cal P}(ax)=P(x).
\label{pax}
\end{equation}
From the above equation and Eqs. (\ref{Pq}) and (\ref{q}) one has
\begin{equation}
{{1}\over{1+rs}}{\cal P}_L({\cal Q})=P_L({\cal Q}_i)=P_L({\mathbb Q}_i).
\label{PQf}
\end{equation}

We have only worked with the variable of interest, in this case ${\cal Q}$. One could, however,
generalize to the energy like variable ${\cal E}$ in order to get the distribution
${\cal P}({\cal E, Q})$ in such a way that
\begin{equation}
{{1}\over{1+rs}}{\cal P}_L({\cal Q})={{1}\over{1+rs}}\int{\cal P}_L({\cal E,\cal Q})d\cal E.
\label{PQE}
\end{equation}
The distribution probability ${\cal P}_L({\cal Q})$ satisfies the scaling relation similar to Eq.(\ref{ssr}), namely
\begin{equation}
{\cal P}_L({\cal Q})=\Lambda_{\cal Q}L^{y_{\cal Q}}
{\cal P}^*_L (\Lambda_{\cal Q}L^{y_{\cal Q}}{\cal Q}),  
\label{P*}
\end{equation}
where ${\cal P}^*$ is the universal distribution. From above, Eq. (\ref{PQf}) 
can be written as
\begin{equation}
P_L({\mathbb Q}_i)={{\Lambda_{\cal Q}L^{y_{\cal Q}}}\over{1+rs}}
{\cal P}_L^*\left({{\Lambda_{\cal Q}L^{y_{\cal Q}}}\over{1+rs}}(Q+sE)\right),
\label{}
\end{equation}
which is of the same form as Eq.(\ref{psr}). Note that the factor $1/(1+rs)$ is
absorbed in the variables when the variance is unit so, analogous to Eq. (\ref{psrn}), one gets
\begin{equation}
{\cal P}_L^*\left((Q_i+sE_i)/\sigma)\right)=\sigma P_L({\mathbb Q}_i),
\label{psr2}
\end{equation}
where $P_L({\mathbb Q}_i)$ is the measured distribution in an MC simulation and 
$\sigma$ is its variance. The temperature, crystal field and $s$ are parameters
that are tuned in order to get the desired universal function. Note, further,
that $\left<Q_i+sE_i\right>$ is different from zero, so one has also to shift the
variable to $(Q_i+sE_i)-\left<Q_i+sE_i\right>$. The distribution ${\cal P^*}$ constructed above provides also a method to determine the first-order transition line, which happens when the shape of ${\cal P^*}$ presents two symmetric peaks at the same height. Of course, its shape will not fit any universal one when the transition is indeed of first order, which prevents to misidentifying a continuous transition. Besides, analytical continuation of first-order line is achieved due to finite-size effects, and critical points should be located accordingly, as we will see in the next section. 

\section{Discrete spin and Lennard-Jones models}

We will present below the use of the mixing fields of the previous section to some discrete spin models as well as to the liquid-gas critical point. 

\subsection{Two-dimensional Spin-$1$  Blume-Capel model}

\begin{figure}[ht]
\includegraphics[clip,angle=-0,width=9cm]{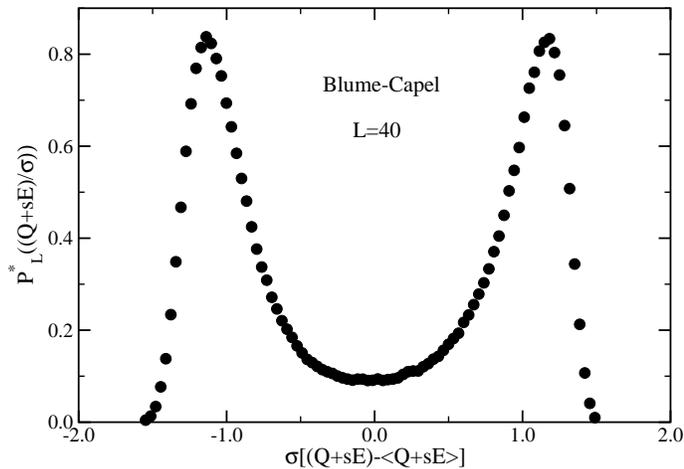} 
\caption{  Probability distribution function of the conjugated extensive variable for the spin-$1$ Blume-Capel model ($L=40$) showing the double peak at the same height.  In this example the system is, according to reference \cite{wn}, at its tricritical point. The errors are smaller than the symbol sizes.}
\label{pdbc}
\end{figure}

\begin{figure}[ht]
\includegraphics[clip,angle=-0,width=9cm]{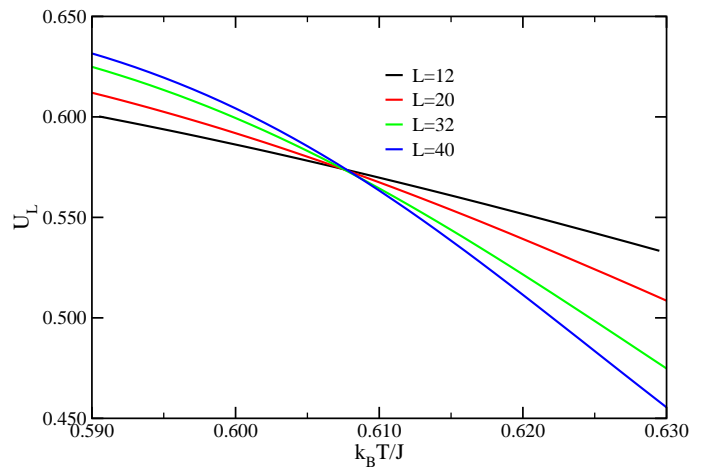} 
\caption{  Fourth-order cumulant $U_L^{\cal Q}$ of the Blume-Capel model along the first-order transition line and its analytic continuation according to reference \cite{wn}. Errors are not shown; only the single histogram result for each lattice is shown.}
\label{uxtbc}
\end{figure}

\begin{figure}[ht]
\includegraphics[clip,angle=0,width=9cm]{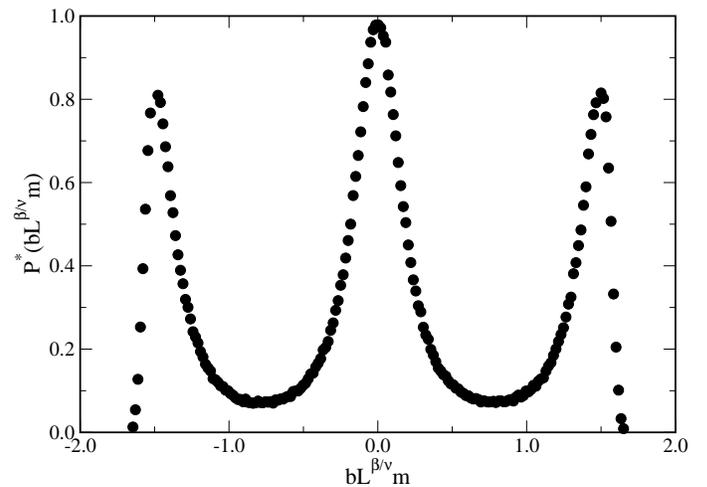} 
\caption{  Probability distribution function of the order parameter for the Blume-Capel model at the tricritical point for $L=40$ according to reference \cite{wn}. The errors are smaller than the symbol sizes.}
\label{pmbc}
\end{figure}

In this case the probability distribution function $P_L(E,Q)$ and $P_L({\mathbb Q})$ can be measured in an MC simulation. One can also easily compute $P_L(Q+sE)$ and its variance $\sigma$ leading to the desired distribution ${\cal P}^*((Q+sE)/\sigma))$ according to Eq. (\ref{psr2}). For a given value of the crystal field $\Delta/J$ one can then tune the temperature $\beta J$ and $s$ in order to have a double peaked function with the same heights. This gives the location of the first-order transition, and from this procedure the corresponding line, as well as its analytic continuation, can be determined. An example of the PDF so obtained is given in Figure \ref{pdbc} for the two-dimensional spin-$1$ model, where $S_i=\pm 1,0$. 
%
In order to get the continuous transition on this first-order line one computes the fourth-order cumulant $U_L^{\cal Q}$ as a function of temperature and seek for their common crossing. This is ilustrated in Figure \ref{uxtbc}, which gives the tricritical temperature and also the tricritical crystal field.  The estimates are  $k_BT_t/J = 0.608(1)$ and $\Delta_t/J=1.9665(3)$, comparable to  $k_BT_t/J = 0.609(4)$ and $\Delta_t/J=1.965(5)$ from reference \cite{zc} and $k_BT_t/J = 0.609(3)$ and $\Delta_t/J=1.966(2)$ from reference \cite{capa}.
%
%
The corresponding order-parameter PDF, which is universal for this class of tricritical phenomena, is shown in Figure \ref{pmbc}. One can clearly see the triple peaks due to the three states of the spin variables. To our knowledge no such distribution for the Blume-Capel model in three dimensions has been published in the literature. However, the shape for the order parameter PDF should be qualitatively similar to that shown in Figure \ref{pmbc}.
%

\subsection{Liquid-Gas Critical Point}

\begin{figure}[ht]
\includegraphics[clip,angle=0,width=8cm]{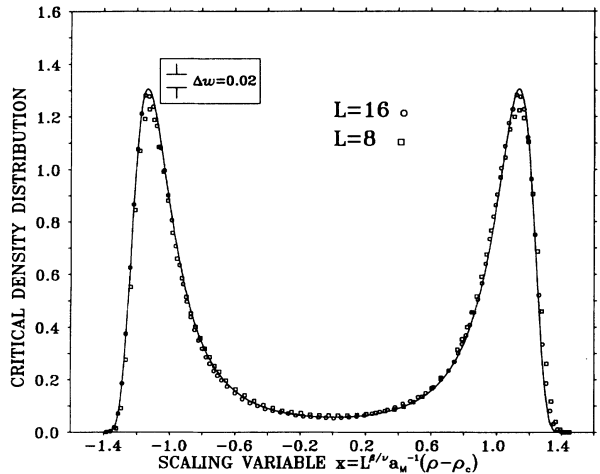} 
\caption{  The fluid probability distribution function of the density $\rho$ at the critical point according to references \cite{bw1,bw2}. The solid line is the order-parameter probability distribution of the Ising universality class and the circles and squares are the data for the Lennard-Jones fluid for lattice sizes $L=16$ and $L=8$, respectively. $\Delta\omega$ is the energy deviation.}
\label{ljf}
\end{figure}

The liquid-gas critical point has been analyzed through MC simulations by Bruce and Wilding in 1992 \cite{bw1,bw2}. Indeed, it was in these seminal papers that they introduced the use of the field mixing  approach in computer simulations for treating asymmetric models and the utility of the universal probability distribution function given by Eq.(\ref{psrn}). In references \cite{bw1,bw2} the two-dimensional Lennard-Jones fluid was studied through Monte-Carlo simulations in the grand-canonical ensemble, from which the first-order transition line (and its analytical continuation) has been determined. The dimensionless energy $\Phi$ of the pair interaction has the form
\begin{equation}
\Phi=4\omega[(\sigma/r)^{12}-(\sigma/r)^6],
\label{lji}
\end{equation}
where $\sigma$ is a constant that sets the interaction range and $\omega$ measures the well depth in units of $k_BT$. 
Differently from the previous case of the Blume-Capel model, the critical point could be determined in this case by just seeking the distribution along the first-order line where a match with that of the Ising universality class is achieved. The results for this model are $4/\omega_c=0.440\pm 0.05,~\mu_c=-2.20\pm 0.04,$ and $\rho_c=0.368\pm 0.003$. As a result, not only the critical point is located but also its universal behavior is determined. Figure \ref{ljf} shows the PDF at criticality compared to that of the spin-$1/2$ Ising model. This is in fact clear evidence, coming from computer simulations, of the fluid-magnet universality, which is independent of any critical exponent value. In addition, an estimate of the ratio of the critical exponents $\beta/\nu=0.125(1)$ is achieved, quite close to the exact value $1/8$. A quantitative comparison between the revised scaling method proposed by Bruce and Wilding \cite{bw1} and the complete scaling procedure by Kim, Fisher, and Orkoulas \cite{kim1} can be found in reference \cite{kim3}.
%
%

Three-dimensional fluid models have also been analyzed according to the present approach. The extension to the three-dimensional Lennard-Jones fluid has been treated by Wilding \cite{wild}, and the decorated lattice gas and a polymer system have been studied by Wilding and M\"uller \cite{wm}. In all cases, the data collapsed onto the three-dimensional PDF shown in Figure \ref{is3d}, providing the location of the critical point as well as the corresponding expected universality class.

 \subsection{Spin-$3/2$  Blume-Capel model}

The Blume-Capel model Hamiltonian (\ref{bch}) for spin-$3/2$ means that $S_i=\pm3/2,\pm1/2$. Besides a second-order phase transition line, the corresponding phase diagram shows a first-order line ending up at a double critical end point \cite{pla2}. Within the present approach, the first-order line (which is in fact a quadruple line where four phases are coexisting) can be determined when the extensive mixing field variables' probability distribution function has peaks of equal heights, and the double critical end point can be located, with a reasonable precision, when this probability distribution matches that given in Figure \ref{is123}. Note that here, differently from the tricritical point of the spin-$1$ Blume-Capel model, one does not need to resort to the cumulant crossings in order to get the continuous transition. For the present model, instead of having a tetracritical point (as should at first sight be expected), one has in fact a double critical end point (two different coexisting critical systems), and it was also possible to convincingly show that this point indeed belongs to the same universality class as the critical ones.

\subsection{Spin-$1$  Baxter-Wu model}

The spin-$1$ Baxter-Wu model is a generalization of the Hamiltonian (\ref{bw}) by considering $S_i=\pm1,0$ and with a crystal field interaction in such a way that the Hamiltonian reads
\begin{equation}
{\cal H} = -J\sum_{i,j,k} S_i S_jS_k-\Delta\sum_{i=1}^NS_i^2.
\label{bw1}
\end{equation}
The phase diagram of the above model is similar in shape to the spin-$1$ Blume-Capel model. However, instead of a triple line ending at a tricritical point one expects a quintuple line (five phases coexisting) ending at a pentacritical point \cite{marluc}. The second-order line should be in the same universality class as the spin-$1/2$ case and the pentacritical point belongs to a different class of universality. 
%
\begin{figure}[ht]
\includegraphics[clip,angle=0,width=9cm]{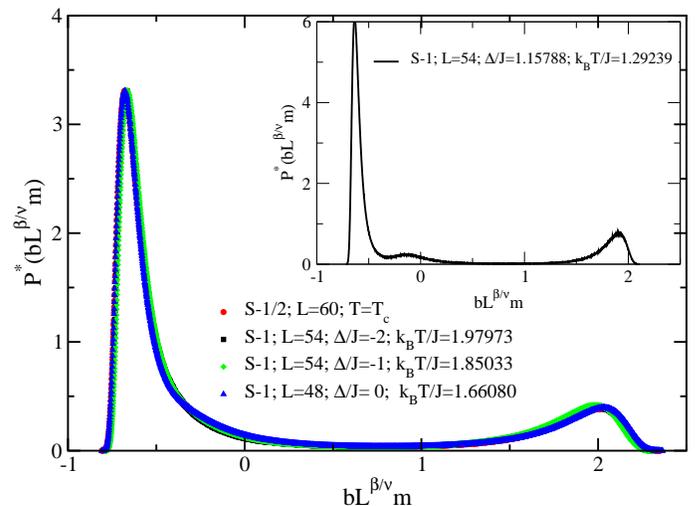} 
\caption{ $P^*$ distribution function of the spin-$1$ Baxter-Wu model. The values of $\Delta/J$ correspond to the continuous transition and are taken at reduced temperatures $k_BT/J$ where the distribution has the same form as that of the spin-$1/2$ model from Figure \ref{bws05} (which is also plotted in this Figure). The inset shows the corresponding distribution function at the pentacritical point. The noise in the distributions give an idea of the errors from the simulations.}
\label{bws1}
\end{figure}
%
Figure \ref{bws1} depicts the order-parameter PDF for values of $\Delta/J$ in the second-order line together with the distribution of the spin-$1/2$ case \cite{marluc2}. One can see that  the continuous transition is in fact in the same universality class, as expected. The inset shows the corresponding distribution function at the multicritical point. As in this case we also have a paramagnetic phase the distribution shows an extra peak for small values of the magnetization $M$.  

\section{Some related models and final remarks}

Other related models have also been studied according to the present approach. For instance, regarding discrete state systems, the isolated critical point of the asymmetric Ising model with two- and three-spin interactions on a triangular lattice, in the presence of an external magnetic field, has been obtained and it has been shown to belong to the Ising universality class \cite{sh1,sh3}. On the other hand, the spin-$1/2$ Ising model on the square lattice with two-body and three-body interactions has been studied\cite{junqi} and very precise distribution functions for much larger systems show the evidence for the  ``universality" of the scaled distribution, including a tricritical point. It is interesting, in this particular case, to make a comparison of the probability distribution function of the order parameter shown in Figure \ref{pmbc}, with the corresponding one for the tricritical point in the Ising model given in Figure 6 of reference \cite{junqi}. While in the former case we have a central peak which is more intense than the two side peaks (because both the $S_i=0$ and disordered states have zero magnetization), for the tricritical point in the Ising model, all three peaks have the same intensity (because there is no $S_i=0$ state). Still regarding discrete models, the $Q$-state Potts model has also been investigated through this technique. The large-$Q$ Potts model in an external field presents an isolated critical point which can also be located through the use of the mixing fields \cite{sh2,sh3}. 

Continuous spin systems and $\phi^4$ models have also been treated by the mixing fields procedure. The PDF of the XY model in two dimensions has been measured \cite{pxy} as well as for the isotropic Heisenberg model. Isolated critical points can be achieved when considering the vectorized Blume-Emery-Griffiths model applied to the phase transition in He$^3$-He$^4$ mixtures \cite{rod,ana}. The  PDF of the $\phi^4$ model has also been studied in reference \cite{nico}.

A different algorithm, namely the grand canonical Monte Carlo method, has been employed with the present scaling technique to study confined lattice homopolymers\cite{panagi}. In a quite recent paper \cite{nathan}, the field mixing revised scaling has been used in order to study the fluid phase behavior of an athermal model of colloids and nano absorbing polymers on a finite lattice. 
More recently, an extension of the probability distribution function for systems in non equilibrium has also been devised in the case of the contact process problem \cite{puline}. 

Thus, it seems that the revised scaling proposed by Rehr and Mermin, associated with the mixing field recipe developed by Wilding and Bruce, is still a useful tool in treating asymmetric magnetic models exhibiting first-order lines and multicritical points. Naturally, a comparison study by applying the complete scaling proposed by Kim and Fisher on magnetic systems would be very welcome.

\acknowledgments{The authors are indebted to Nigel Wilding for drawing their attention to the complete scaling procedure, for sending his original files regarding Figures \ref{pdbc}, \ref{uxtbc}, and \ref{pmbc}, for several suggestions and for a critical reading of the manuscript. Special thanks to Shan-Ho Tsai for fruitful private conversations are also addressed. R.S.T. Freire is also acknowledged for a critical reading of the manuscript. Financial support comes from the  brazilian Agencies CNPq, CAPES, FAPEMIG, FAPEMAT, and SBF. The authors would like to thank Prof. David Landau for several suggestions regarding the manuscript and for kind hospitality at the Center for Simulational Physics (CSP) at University of Georgia (UGA, Georgia, USA), where part of this work has been done. JAP thanks the Sociedade Brasileira de F\'isica (SBF) - American Physical Society (APS)  Professorship/Lecturership 2011 program where part of the subject of this paper has been presented as informal lunch seminars at CSP. PHLM thanks CAPES and UFMT for the post-doctoral scholarship at CSP-UGA.}


\begin{thebibliography}{00}
\bibitem{text} The revised scaling is not able to describe the Yang-Yang anomaly present in highly asymmetric fluids. According to Yang and Yang \cite{yy} the second derivative, ${d^2\mu_\sigma}\over{dT^2}$, where $\mu_\sigma(T)$ is the chemical potential along the coexisting line for $T<T_c$, diverges, as does the specific heat, for $T\rightarrow T_{c-}$. But, from the scaling forms (\ref{sg}) and (\ref{asg}) the chemical potential is an analytic function of temperature. The complete scaling procedure \cite{kim1}, where the pressure is also mixed as the other $t$ and $\mu$ fields, is able to correctly describe the Yang-Yang anomaly. However, in addition to the $1-\alpha$ divergence in the diameter density, it predicts an extra leading contribution on the diameter as $t^{2\beta}$. Unfortunately, experimental results are not yet conclusive in order to clarify these important leading singularities and still remain an open question \cite{kim3}.
\end{thebibliography}
\end{document}